\documentstyle[epsf,epsbox,mbf,wrapft]{aipproc}

\begin{document}
\title{The Values Distribution in a Competing Shares Financial Market
Model}

\author{A. Ponzi and Y. Aizawa} 
\address{Department of Applied
Physics, Waseda University, 3-4-1 Okubo, Shinjuku-ku, Tokyo 169-0072,
Japan}

%\lefthead{LEFT head}
%\righthead{RIGHT head}
\maketitle

\begin{abstract}
We present our competing shares financial market model and describe
it's behaviour by numerical simulation. We show that in the critical
region the distribution avalanches of the market value as defined in
this model has a power-law distribution with exponent around $2.3$.
In this region the price returns distribution is truncated Levy
stable.
\end{abstract}

\section*{Introduction and Model}
Recent studies{\cite{MS1}} of the S\&P500 financial market index have
shown that financial markets are not appropriately described by the
Efficient Market Hypothesis which predicts a Gaussian type behaviour
of the price changes time series.{\cite{pet1}} In this paper we will
present a reinterpretation of our toy financial market model which was
presented in {\cite{PA2}}. That model was a trader based model, here
however we re-present that model as a share based system. That model
was a simplified mean-field version of our neural-network type
financial market model presented in {\cite{PA1}}. Although this model
was developed independently, it is a variant of the Minority
Game{\cite{CZ}}, with however some fundamental differences. We explain
this model here as follows.\\
\indent There are $N$ shares labeled $i=1,..,N$ (or commodities etc)
in a competetive market. Each share at time $t$ can be in one of two
states $s_{i}(t)$ which classify the majority ownership crowd of the
share $i$, $s_{i}(t)=1$ `bull', $s_{i}(t)=-1$ `bear'. Bulls have
bought the share, hoping to sell it later at a profit, bears have sold
the share hoping to buy it back later at a profit. The price returns
of each share are given by, 
\begin{equation}
\Delta p_{i}(t)=\frac{N}{2}s_{i}(t).
\label{eqn:price}
\end{equation}
that is like excess demand, with price returns change being
proportional to the `size' of the market $N$. Each share also has a
{\it generally perceived value} $V_{i}(t)$, which ranks the shares
$i$. High valued shares have recognized long term price trend with
therefore {\it stuck}, low volatility, states $s_{i}(t)$. Low valued
shares on the other hand are short term {\it risky} states with {\it
high volatility}. The state update dynamic is Darwinian competition;
we choose two shares at random, say $a$ and $b$ with
$V_{a}(t)>V_{b}(t)$, then,
\begin{equation}
s_{i}(t+1)=\left\{\begin{array}{ll} s_{i}(t), & (a)\\
-s_{i}(t), & (b)
\end{array}
\right.
\label{eqn:spinup}
\end{equation}
The shares therefore define a jammed flow of investors, long term
investors up the values, short term investors down the values. The
high valued shares become stuck in bull(bear) states because nobody
wants to sell(buy) them.  In other words investors looking for long
term investments look for the low volatility high valued shares, and
therefore reinforce these shares price trends by trying to
take the corresponding position on them. Low valued shares on the
other hand are left in fluctuating states without any well-defined
price trend and therefore have high-volatilty since nobody knows which
way they are going. In this model therefore, both price trend and high
volatilty self-reinforce themselves. Hence the spin update dynamic
Eq.\ref{eqn:spinup}, which basically says that high valued are low
volatility and low valued high volatility.\\
\indent
In reality a shares value is a complex function of individual company
news (affecting each share separately), macroscropic news(affecting
all shares), personal traits, the weather etc. However in this model
we model only the speculative behaviour of traders. The shares values
are therefore defined as follows,
\begin{equation}
V_{i}(t+1)=V_{i}(t)-\frac{1}{2}\Delta s_{i}(t)G(t)
-(1-\frac{1}{2}|\Delta s_{i}(t)|)c
\label{eqn:values}
\end{equation}
where $\Delta s_{i}(t)=s_{i}(t+1)-s_{i}(t)$ and our variable
`groupthink' $G(t)$ is the overall market (macroeconomic) state,$
G(t)=\frac{1}{N}\sum_{1}^{N}s_{i}(t)=\frac{2\Delta p(t)}{N}$ and
$p(t)$ is the share index price return. Therefore the stuck shares (a)
have slowly decreasing value $(c>0)$ due to rise(fall) in the price of
bull(bear) states. Risky shares (b) however, which are the domain of
speculators and gamblers, increase in value when they move into the
minority state of the {\it overall} market. Here we are assuming that
the macroeconomic bull/bear state $G(t)$ is coupled to the individual
share states $s_{i}(t)$ in the same way as the usual Minority Game,
except on a larger scale.\\
\indent
This values update rule Eq.\ref{eqn:values} implies that the two main
observables for investors picking between shares are the individual
shares volatilities, which are related to the first term, and the
individual shares price trends, which are related to the second
term. Indeed long term investors may `play' the observed price trends,
while short term speculators may `play' the volatility.\\
\indent
Now we generalise the Darwinian evolution dynamic by defining a
probability for dynamic $a$ or $b$ update in Eq.\ref{eqn:spinup},i.e.
\begin{equation}
Probability(a)=1-Probability(b)=\frac{1}{1+e^{-2\beta
v_{i}(t)}}
\label{eqn:prob}
\end{equation}
where $\beta$ is a kind of inverse temperature parameter, and
$v_{i}(t)=V_{i}(t)-V(t)$ is the {\it relative} value, with
$V(t)=\frac{1}{N}\sum_{1}^{N}V_{i}(t)$ the mean-value (market-value). 
This implies that indeed high valued shares are low volatility and low
valued shares high volatility, with a volatility gradient which
depends on the parameter $\beta$. This simply replaces our dynamic of
randomly choosing two shares $(a)$ and $(b)$ and comparing the values. 
This evolution dynamic is similar to co-evolution on coupled fitness
landscapes{\cite{Kauff1,BS,SM1}}, since the relative values $v_{i}(t)$
can change due to changes in $V_{i}(t)$ or in a co-evolutionary sense
by changes in $V(t)$.\\
\begin{figure}[htp] % fig 1xn25ts.eps
\centerline{\epsfile{file=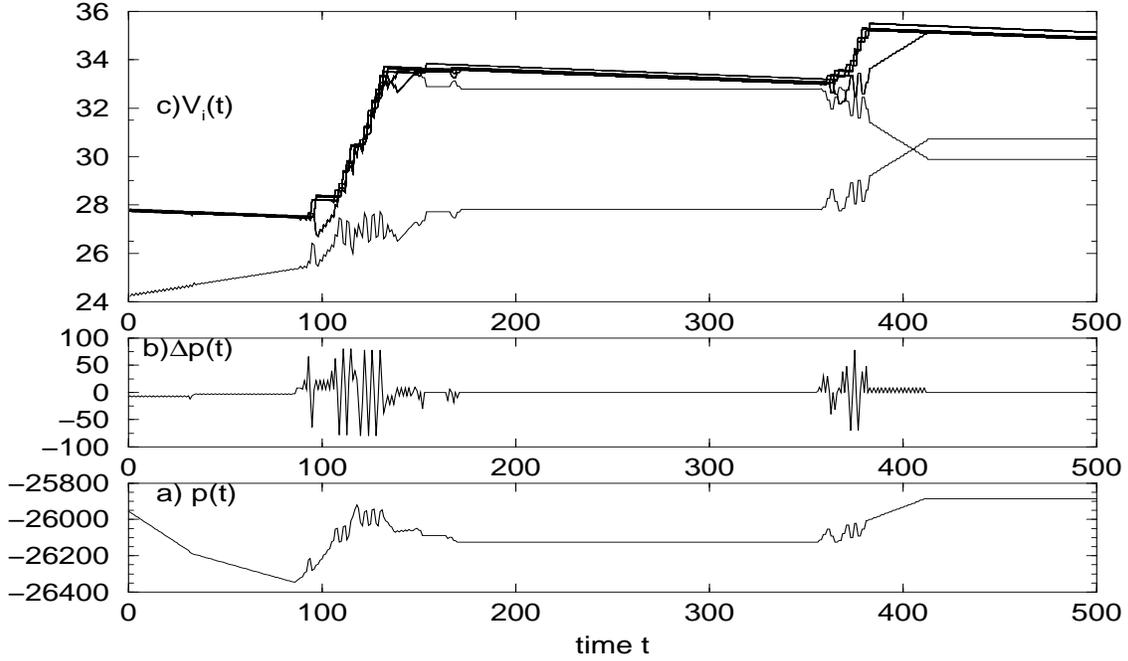,height=9cm,width=15cm}}
%\vspace{10pt}
\caption{$c=0.003$, $N=80$ $\beta=\infty$ time series. a)Index price returns $p(t)$, b)Index price returns changes $\Delta p(t)$, c)Values $V_{i}(t)$}
\label{fig1}
\end{figure}
\indent Our dynamic is then that we first calculate $G(t)$ and $V(t)$
and then we update all shares $V_{i}(t)$ and $s_{i}(t)$ according to
probability given by Eq.\ref{eqn:prob}. By putting $\beta=\infty$ we
obtain the deterministic system where if $V_{i}(t)>V(t)$ then $(a)$
and otherwise $(b)$. For the initial conditions for which we choose
$V_{i}(0)\in[-1,1]$ uniformly randomly and $s_{i}(0)=\pm1$ uniformly
randomly.\\ 
\indent A more general version of this model defines the
price changes by extending the definition Eq.\ref{eqn:price}, to,
\begin{equation}
\Delta p_{i}(t)=\frac{N}{2}s_{i}(t)v_{i}(t)
\label{eqn:newprice}
\end{equation}
where $v_{i}(t)$ are the relative values, which we believe is more
realistic than the simple definition Eq.\ref{eqn:price}. However in
this paper we confine to Eq.\ref{eqn:price}, since the study of
Eq.\ref{eqn:newprice} is still in progress.
\indent
This model resembles the MG. However it is different in two
fundamental ways. 1) We only apply the minority rule when the state
changes, that is the amount a shares value is updated according to the
minority rule is proportional to it's volatility across any time
period. 2) {\it There is no strategy space.} We simply map straight
from the values to the state update rule. 
\section*{Results}
First we describe time series behaviour of the deterministic
system\cite{PA2,PA4}. Shown in Fig.\ref{fig1} are time series for
$\beta=\infty$. For small $c$ and large $N$ the values $V_{i}(t)$
shows `Punctuated Equilibrium'{\cite{GE}} type behaviour reminiscent
of ecodynamics, i.e. avalanches and stasis periods.

\begin{wrapfigure}{r}{7 cm}
\epsfxsize=7 cm 
%or \epsfysize = 9 cm
\centerline{\epsfbox{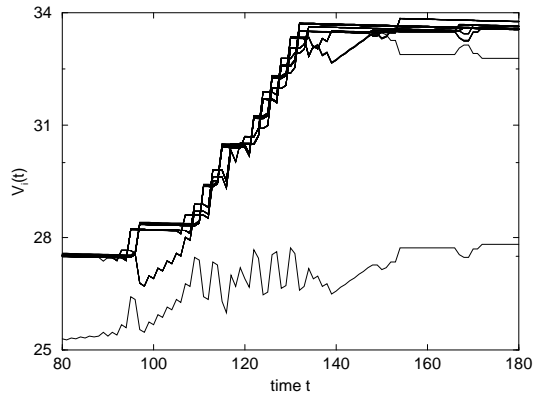}}
\caption{Detail of Values Time series from Fig.1}
\label{fig2}
\end{wrapfigure}

In fact the avalanches are, to varying degrees, bursts of partial
global synchronization where most of the share values $V_{i}(t)$
cluster into 2 or more groups and interleave the mean-value, growing
rapidly in a co-operative way (see Fig.\ref{fig2}), and represent an
unstable attractor for the system. A few shares are left out and
because this synchronization behaviour rapidly increases the values
deviation $d(t)=\frac{1}{N}\sum_{1}^{N}v_{i}(t)$ we call these bursts
`flights to quality', like the phenomena which occur in real financial
markets from time to time, such as the Russian Crisis in 1998. 

\begin{wrapfigure}{r}{7 cm}
\epsfxsize=7 cm 
%or \epsfysize = 9 cm
\centerline{\epsfbox{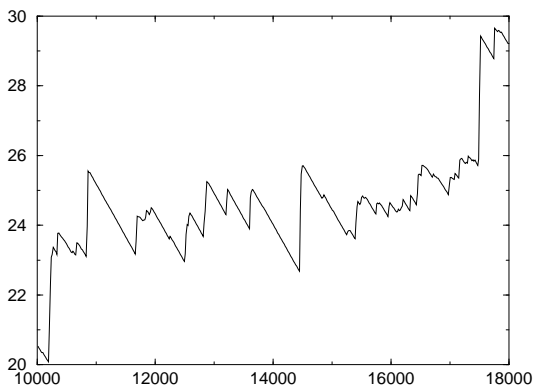}}
\caption{$N=80$, $c=0.003$, $\beta=\infty$, Mean-value $V(t)$ time
series, (longer time length)}
\label{banktot}
\end{wrapfigure}

They
occur due to the driving term $c$ decreasing $d(t)$ and eventually
causing one value to cross the mean-value thereby possibly starting an
avalanche, and occur at small $d(t)$ when the market is susceptible to
fluctuation (here internally generated by chaos).
Indeed at times of small $d(t)$ nobody knows what is best to
invest in and panics and stampedes may occur due to wild speculation
(irrespective of the actual information contained in the item of news
which hit the market). These bursts show up as oscillations
(Fig.\ref{fig2}(b)) in the price changes time series $\Delta p$, we
call them `market rollercoasters'. In fact the slightly oscillatory
nature of financial time series has been noted{\cite{pet1,mand1}. The
price time series shown (Fig.\ref{fig2}(c)) is very reminiscent of the
technical-analysts `double tops' and other recognized formations. In
this view a market will interpret rumour as good or bad dependent on
the current state $G(t)$, rather than on the value of the news itself,
such as when the meeting of James Baker and Tariq Aziz before the Gulf
War went on a little too long and caused a market rollercoaster. In
fact when a market is over-bought (most people in bull states) the
only way it can go is into sell mode and it is not so surprising that
momentum may carry it into an over-sold state.\\

\begin{figure}[htp] 
\centerline{\epsfile{file=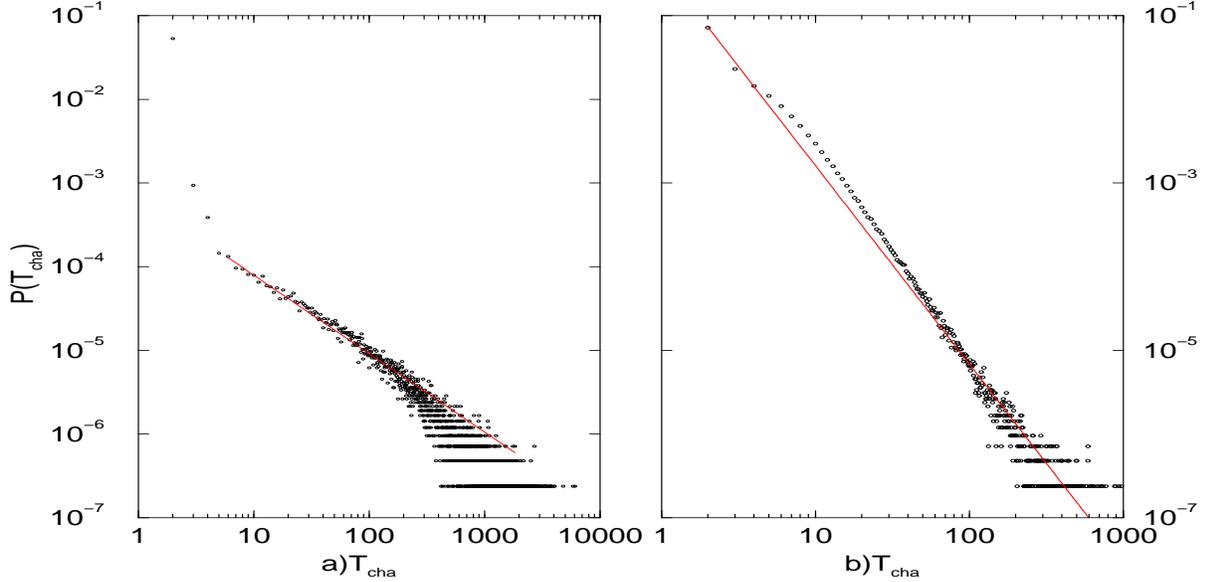,height=8cm,width=16cm}}
%\vspace{10pt}
\caption{Probability distribution of lengths of periods of stasis,
$T_{cha}$. a)Slope $-0.93\pm0.01$. $c=0.001$, $N=200$, $\beta=\infty$,
b)Slope $-2.36\pm0.03$,$c=0.001$, $N=200$, $\beta=80$.}
\label{fig3}
\end{figure}
\indent The most fundamental variables in our model are the values
 rather than the prices and so we study their behaviour. A mean-values
 time series is shown in Fig.\ref{banktot}. In{\cite{PA2}} we showed
 that for small $c$ this system is critical in the sense that
 avalanches, defined as size of changes in $V(t)$, have a power law
 distribution with a peak due to the almost periodic state. Here we
 show that the stasis periods also show such behaviour. Shown in
 Fig.\ref{fig3}(a) is the distribution of $T_{cha}$ where $T_{cha}$ is
 the time between successive events $\Delta V(t)>0$, they are
 normalized by dividing by the whole time series length
 $T_{len}=4\times10^6$, after discarding a long transient. Only one
 time series is included in this figure, so periods of stasis of all
 sizes (limited by the system size) are always present. The slope of
 the line is very near 1, which is a good fit for the longer $T_{cha}$
 however at small $T_{cha}$ there is an increase in probability due to
 time spent in the almost periodic states. This shows that total
 market value itself has a Punctuated Equilibrium time behaviour,
 independent of `external' information. External information may
 however initiate an avalanche itself as mentioned above. \\
\indent In fact more realistically we may study the stochastic system
defined by setting $\beta\neq\infty$. We found that as we change
$\beta$ we see a phase transition, similar to that seen in the more
general MG{\cite{CZ}}, of which this is model is a restricted
version{\cite{PA3}}. Furthermore in the transition region the price
returns distribution defined by Eq.\ref{eqn:price} shows a Levy
distribution\cite{LP1,LP2} for the central values, while for values
after about four standard deviations from the mean there is a drop-off
of the probability.{\cite{PA3}} Time series in this region at
$\beta=80$ are shown in Fig.\ref{ts80}. 
\begin{wrapfigure}{r}{7 cm}
\epsfxsize=7 cm 
%or \epsfysize = 9 cm
\centerline{\epsfbox{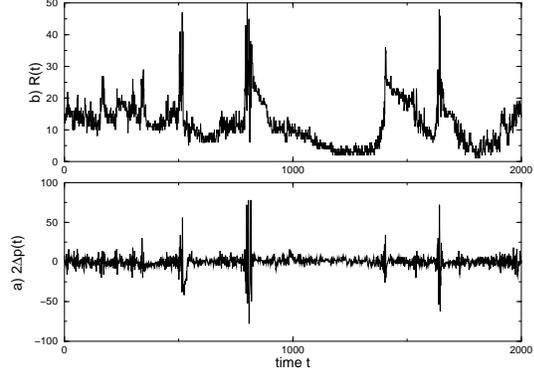}}
\caption{$N=200$, $\beta=80$, $c=0.001$ a)Price changes time series,
b)Trading volume $R(t)$ defined by the amount of spins which flip.}
\label{ts80}
\end{wrapfigure}
The parameter characterising
the Levy distribution we found was about 1.5 which is very similar to
the actual distribution for the $S\&P500$ measured by Mantegna and
Stanley{\cite{MS1}}. Here we show that the corresponding mean-value
changes $\Delta V$ distributions at $\beta=80$ still show a good
scaling behaviour, where however the exponents have changed. Shown in
Fig.\ref{fig3}(b) is the $\beta=80$ $T_{cha}$ distribution and in
Fig.\ref{fig:btcha} the distribution of changes $\Delta V$, divided
into both positive and negative contributions. Again only one time
series of length $T_{len}=4\times10^6$ was included. The exponents
have changed from near 1 for the deterministic system to around $-2.3$
as shown in the figure captions.
\begin{figure}[htp] 
\centerline{\epsfile{file=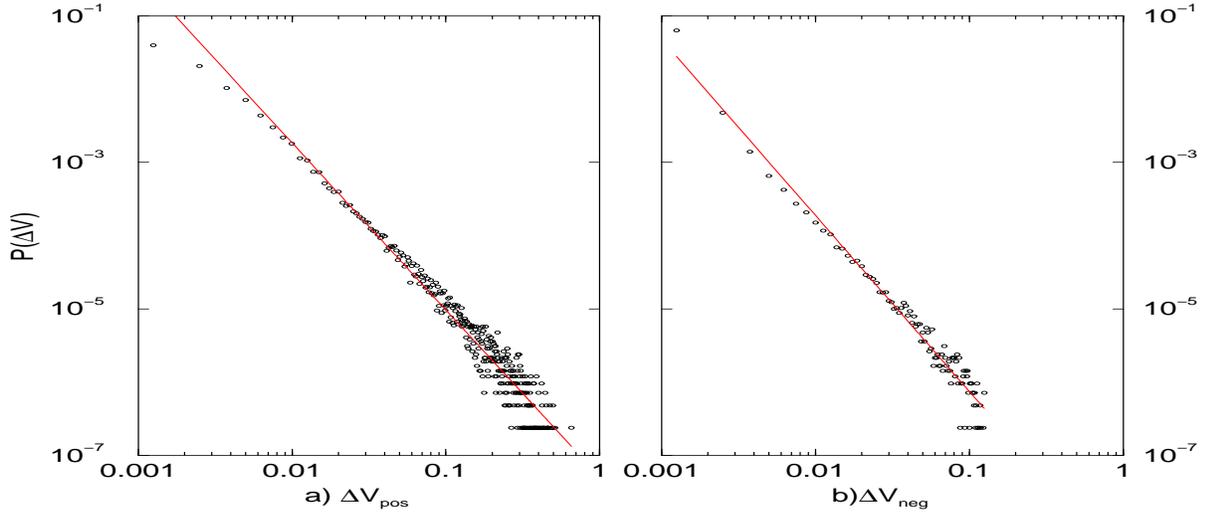,height=7cm,width=16cm}}
%\vspace{10pt}
\caption{Probability distribution of avalanche sizes $\Delta V$,
$c=0.001$, $N=200$, $\beta=80$, a)Positive changes, Slope
$-2.28\pm0.03$. $c=0.001$, $N=200$, $\beta=80$, b)Negative changes
slope $-2.40\pm0.05$}
\label{fig:btcha}
\end{figure}
Recent work{\cite{GOP}} has studied price returns distributions for
the $S\&P500$ and found that while the central region of the
distribution may be characterised by a Levy stable distribution with
parameter between 1.35-1.8 the tails fall off with an exponent of
around 3. We are at present studying the behaviour of the tails of the
price returns distributions for both Eq.\ref{eqn:price} and
Eq.\ref{eqn:newprice}, which may be more appropriate, to see if they
give the correct results.
\section*{Discussion}
This model may seem naive but it seems to well-reproduce many observed
characteristics of financial market time series, including qualitative
features. Furthermore it is built on fairly simple assumptions and can
explain results that other models based on a single share cannot, for
example the fact that usually all shares crash together, and that all
individual shares have similar distributions.{\cite{Pler}}. This
should be investigated further. This model predicts many interesting
relationships between share prices, especially the values deviation
$d(t)$. To what extent the behaviour of $d(t)$ corresponds with
reality is under investigation.

\end{document}